\documentclass[12pt]{iopart}

\usepackage{graphicx}

\begin{document}

\title[Optimal cloning of single photon polarization
by coherent feedback]{Optimal cloning of single photon
polarization by\\ coherent feedback of beam splitter losses}

\author{Holger F. Hofmann \dag\ and Toshiki Ide \ddag\
}

\address{\dag\ Graduate School of Advanced Sciences of Matter, Hiroshima University,
Kagamiyama 1-3-1, Higashi Hiroshima 739-8530, Japan}

\address{\ddag\ Okayama Institute for Quantum Physics,\\
1-9-1 Kyoyama, Okayama City, Okayama, 700-0015, Japan}

\begin{abstract}
Light fields can be amplified by measuring the field
amplitude reflected at a beam splitter of reflectivity $R$
and adding a coherent amplitude proportional to the measurement
result to the transmitted field. By applying the quantum
optical realization of this amplification scheme to
single photon inputs, it is possible to clone the
polarization states of photons. We show that optimal cloning
of single photon polarization is possible when the gain factor
of the amplification is equal to $1/\sqrt{1-R}$.
\end{abstract}


\ead{h.hofmann@osa.org}

\maketitle

\section{Introduction: quantum cloning and field amplification}

One of the fundamental features of quantum information
is that it is impossible to generate perfect copies
(or ``clones'') of an unknown quantum state input \cite{Woo82}.
This no-cloning theorem is particularly interesting
in the light of the wave-particle dualism of optics, since
the amplitude of a classical wave can be copied perfectly
by any classical amplification process.
Soon after the first formulation of the no-cloning theorem,
it was pointed out that perfect cloning by phase sensitive
optical amplification is prevented by the unavoidable spontaneous
emission in such processes \cite{Mil82,Man83}.
However, it was later found that stimulated emission is in fact
an optimal approximation to perfect quantum cloning \cite{Sim00}.
This insight was quickly followed by the first
experimental realizations of optical quantum cloning using
parametric optical amplification \cite{Mar00,Lam02,Mar02,Fas02}.
Recently, it has also been discovered that the bunching
properties of light fields can be used to obtain optimal clones
by post-selecting the output of a beam splitter \cite{Irv04}.
In general, optical cloning methods thus exploit the
natural wave-particle dualism of light to clone the quantum
coherence of photons by manipulating the (classical) optical
coherence of the light field.

A more direct way to access the field properties of photons
is to measure the quadrature components $\hat{x}$ and $\hat{y}$
of the complex field amplitude, $\hat{a}=\hat{x}+i \hat{y}$.
As demonstrated by a number of experimental results
\cite{Lvo01,Lvo02a,Lvo02b,Zav04a,Zav04b}, such measurements provide
quantum mechanically precise information on the coherent
field properties associated with photon number states.
It seems obvious that this method can also be used to measure
the polarization state of a photon, since the
polarization of light is completely described by the two
complex amplitudes $\hat{a}_H$ and $\hat{a}_V$ of a pair
of orthogonal polarizations $H$ and $V$.
For a single photon input, the measurement of the
two complex amplitudes $\hat{a}_H$ and $\hat{a}_V$ by
homodyne detection is indeed equivalent to a quantum
mechanically precise detection of the photon in
the polarization defined by the measurement results
obtained for the amplitudes.
A particularly simple cloning scheme could thus be realized
by measuring the complex amplitudes of the input photon
and modulating a coherent laser beam to emit multiple photons
with the same polarization amplitudes.

However, homodyne detection can also be applied to fields
of unknown photon number. It is then possible to
obtain partial information about the polarization of a photon
by ``dividing'' the one photon input at a beam splitter
of reflectivity $R$ and measuring only the reflected fraction
of the light. The resulting losses caused by the reduction
of the transmitted amplitude by a factor of $\sqrt{1-R}$
can be compensated by adding a coherent laser amplitude
proportional to the measurement results for the coherent
amplitudes $\hat{a}_H$ and $\hat{a}_V$ \cite{Hof01}.
It is also possible to over compensate the losses to achieve
an amplification of the field variables.
In fact, it has been shown that this kind of over compensation
can be used to achieve the noiseless amplification of a single
quadrature component of the light field \cite{Lam97}, and
the application of this scheme to the continuous variable
cloning of Gaussian states has recently been demonstrated
\cite{And05,Oli06}. It is thus clear that a minimal noise
amplification of the light field can be achieved by a finite
resolution measurement of the amplitude and an appropriate
coherent feedback.

In the following, it is shown that the kind of
optical amplification used to clone continuous variable field
states of a single mode in \cite{And05} can also be used to
clone the polarization state of a single photon input.
By using an optimized gain factor of $1/\sqrt{1-R}$ to
minimize the noise effects in the amplification,
it is possible to achieve optimal cloning of the
qubit encoded in the single photon polarization.
Interestingly, this kind of cloning process does not require
any optical non-linearity to achieve the desired transfer
of polarization from the one photon input to the multi-photon
output. Instead, the phase information needed to clone a quantum
coherent state is obtained explicitly in the form of
an optical measurement of field coherence, and the amplification
is performed by adding the desired light field amplitude
using linear interference between the transmitted light
and an appropriately modulated strong laser field.
During this process, photon number is not preserved, and the
quantum information is transferred from the input photon to
the output photon by quantum coherences between states of
different photon number. It is thus possible to exploit fundamental
aspects of the wave-particle dualism in order to manipulate
the discrete polarization statistics of photons through continuous
variable operations.

\begin{figure}[ht]
\begin{picture}(400,320)
\thicklines

\put(200,60){\line(0,1){60}}
\put(202,58){\makebox(85,12){Beam splitter,}}
\put(202,46){\makebox(85,12){Reflectivity $R$}}

\put(190,105){\line(-1,1){45}}
\put(185,100){\line(-1,1){45}}
\put(140,150){\line(0,-1){10}}
\put(140,150){\line(1,0){10}}

\put(210,105){\line(1,1){45}}
\put(215,100){\line(1,1){45}}
\put(260,150){\line(0,-1){10}}
\put(260,150){\line(-1,0){10}}

\put(150,45){\line(1,1){35}}
\put(155,40){\line(1,1){35}}
\put(190,80){\line(0,-1){10}}
\put(190,80){\line(-1,0){10}}

\put(90,25){\makebox(40,15){1 photon input,}}
\put(90,10){\makebox(40,15){polarization state
$\mid \psi_{\mbox{in}}\rangle$}}

\put(130,130){\line(0,1){60}}
\put(40,126){\makebox(85,15){Beam splitter,}}
\put(40,111){\makebox(90,15){Reflectivity 1/2}}

\put(120,175){\line(-1,1){35}}
\put(115,170){\line(-1,1){35}}
\put(80,210){\line(0,-1){10}}
\put(80,210){\line(1,0){10}}
\put(50,215){\makebox(40,20){$x_H/\sqrt{2}$,
$x_V/\sqrt{2}$}}
\put(140,175){\line(1,1){35}}
\put(145,170){\line(1,1){35}}
\put(180,210){\line(0,-1){10}}
\put(180,210){\line(-1,0){10}}
\put(170,215){\makebox(40,20){$y_H/\sqrt{2}$, $y_V/\sqrt{2}$}}

\put(70,245){\framebox(120,60){}}
\put(80,285){\makebox(100,15){Measurement of}}
\put(80,265){\makebox(100,15){$\beta_H=x_H+i y_H$}}
\put(80,250){\makebox(100,15){$\beta_V=x_V+i y_V$}}

\bezier{400}(190,280)(240,280)(280,200)
\put(280,200){\line(0,1){17}}
\put(280,200){\line(-3,2){15}}

\put(255,155){\framebox(60,40){$\hat{D}(f_R \vec{\beta})$}}

\put(310,205){\line(1,1){25}}
\put(315,200){\line(1,1){25}}
\put(340,230){\line(0,-1){10}}
\put(340,230){\line(-1,0){10}}

\put(320,250){\makebox(40,15){$N$-photon output,}}
\put(320,235){\makebox(40,15){polarization
state $\hat{\rho}_N$}}
\end{picture}
\setlength{\unitlength}{1pt}
\caption{\label{fig1} Schematic setup of the optimal cloning machine.
The one photon input state $\mid \psi_{\mbox{in}} \rangle$ is
split at a beam splitter of reflectivity $R$. The reflected part
is split once more to allow the simultaneous uncertainty limited
measurement of the four quadrature components $\hat{x}_H$,
$\hat{x}_V$,$\hat{y}_H$ and $\hat{y}_V$ by homodyne detection.
The measurement result is then transmitted to an
optical modulation setup that displaces the transmitted
field amplitudes by a feedback of $f_R$ times the measured
amplitudes.}
\end{figure}
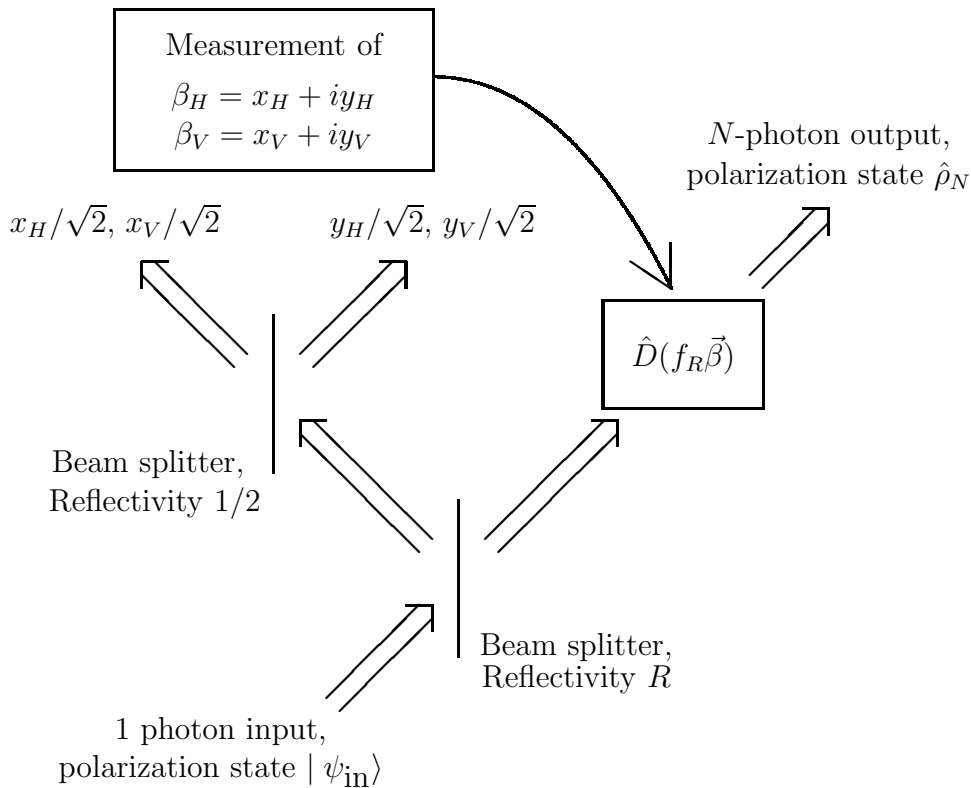


\section{Theory of homodyne detection for a single
photon input}

Fig. \ref{fig1} shows the schematic setup of the proposed
optimal cloning machine. The center piece is the beam
splitter of reflectivity $R$ that splits the single photon
input into two fields. The quadrature components of the
reflected field are then measured by homodyne detection,
and a coherent feedback is applied to displace the field
amplitude of the transmitted field by $f_R$ times the
measurement result $\vec{\beta}=(\beta_H,\beta_V)$.
In principle, this setup corresponds to the setups
for noiseless amplification \cite{Lam97}, compensation
of beam splitter losses \cite{Hof01}, and Gaussian state cloning
\cite{And05}. However, in order to handle polarization states,
the present setup has to amplify a total of four quadrature
components, corresponding to a pair of two orthogonal
polarization modes $\hat{a}_H$ and $\hat{a}_V$.
For the following discussion, it will be most convenient to
define the input state in terms of the creation operators
$\hat{a}_H^\dagger$ and $\hat{a}_V^\dagger$ of these two
polarization modes, since these operators also represent the
complex conjugate field amplitudes permitting a particularly
simple description of linear optics operations on the photon number
states. The unknown polarization state of the input photon can then
be written in the photon number basis of the two mode field as
\begin{equation}
\mid \psi_{\mbox{in}} \rangle_{a} =
c_H \mid 0;1 \rangle_{HV} + c_V \mid 1;0 \rangle_{HV}
=\left( c_H \hat{a}_H^\dagger + c_V \hat{a}_V^\dagger\right)
\mid 0;0 \rangle_{HV}.
\end{equation}
The quantum information encoded in this state is expressed
by the probability amplitudes $c_H$ and $c_V$ of the
horizontally ($H$) and vertically ($V$) polarized one
photon states.

The effect of the beam splitter on this state can be
obtained by transforming the input modes
into a coherent superposition of transmitted modes
$\hat{a}_i$ and reflected modes $\hat{b}_i$,
\begin{equation}
\label{eq:split}
\fl
\hat{U}_{\mbox{\tiny BS}}\!
\mid \! \psi_{\mbox{in}} \rangle_{a} \otimes\!
\mid \! 0;0 \rangle_{b} =
\left(\sqrt{1-R}( c_H \hat{a}_H^\dagger +
c_V \hat{a}_V^\dagger) + \sqrt{R}(c_H \hat{b}_H^\dagger +
c_V \hat{b}_V^\dagger \right)
\mid \! 0;0\rangle_{a} \otimes\! \mid \! 0;0 \rangle_{b}.
\end{equation}
The quantum information is now distributed between the
reflected modes and the transmitted modes in the form of an
entanglement between the modes.

As indicated by fig. \ref{fig1}, the reflected mode is then
split into equal parts at a beam splitter of reflectivity
$1/2$, and polarization sensitive homodyne detection
is applied to both parts to obtain the two quadrature
components of the complex amplitudes
$\beta_H=x_H+i y_H$ and $\beta_V=x_V+i y_V$.
The continuous variable measurement of the reflected beam then
projects the field in the transmitted beam into a superposition
of vacuum and one photon components, resulting in a corresponding
superposition in the transmitted modes. Note that experimentally,
this is similar to the recently demonstrated preparation
of photonic qubits by field quadrature noise measurements \cite{Bab04}.
The main technical difference of our procedure is that we use an
additional beam splitter to achieve an uncertainty limited simultaneous
measurement of both quadrature components, $x_{H/V}$ and $y_{H/V}$.
In terms of quantum measurement theory, this kind of measurement
projects a general input state in the reflected modes $\hat{b}_i$
onto a coherent field state $\mid \! \beta_H; \beta_V \rangle_{HV}$
where the amplitudes $\beta_H$ and $\beta_V$ define the
measurement outcome \cite{Hof01}.
The properly normalized positive operator valued measure
of this measurement reads
\begin{equation}
\mid P({\vec{\beta}}) \rangle = \frac{1}{\pi}
\mid \beta_H; \beta_V \rangle_{HV}
\hspace{0.5cm} \mbox{with} \hspace{0.5cm}
\int d^4{\vec{\beta}} \mid P({\vec{\beta}}) \rangle \langle
P({\vec{\beta}}) \mid = \hat{1}.
\end{equation}
The conditional output state of the transmitted field
modes $\hat{a}_i$ after the measurement is then given by
\begin{eqnarray}
\label{eq:cond}
\fl
\sqrt{p({\vec{\beta}})} \mid \psi({\vec{\beta}}) \rangle_{a}
&=& \frac{1}{\pi} _{b}\langle \beta_H; \beta_V \mid
\hat{U}_{\mbox{BS}} \mid \psi_{\mbox{in}} \rangle.
\nonumber
\\
&=& \frac{1}{\pi}
\mbox{e}^{-|\vec{\beta}|^2/2}
\left(\sqrt{1-R} ( c_H \hat{a}_H^\dagger +
c_V \hat{a}_V^\dagger) + \sqrt{R} (\beta_H^* c_H + \beta_V^* c_V) \right) \mid 0;0\rangle_{a},
\end{eqnarray}
where $p({\vec{\beta}})$ is the probability of obtaining the
measurement result ${\vec{\beta}}$. This state is a coherent superposition
of a single photon state with the original input polarization and a
vacuum component, where the quantum coherence between the vacuum and the
single photon states is defined by the relation between the input state
amplitudes $c_H, c_V$ and the measurement results $\beta_H, \beta_V$.

\section{Coherent feedback and optimized gain}
It is now possible to modify the output state by coherently
adding field amplitudes of $f \beta_H$ and $f \beta_V$ to
the polarization components of the output field.
As demonstrated in \cite{Lam97,And05}, this kind of field
addition can be achieved in a straightforward manner by interfering the output field and an appropriately modulated laser beam
at a highly reflective beam splitter. However, it may be worth
noting that these experiments were performed in continuous wave
operation, while a single photon state must be defined in terms
of a finite pulse shape \cite{Lvo01,Lvo02a,Lvo02b,Zav04a,Zav04b}.
In our case, the physical system is thus defined by a single
optical pulse, and the conditional displacement must be timed
to act on the same pulse from which the measurement data was
obtained. This identity of the measured pulse with the output
pulse is also the reason why we refer to the conditional displacement
as a feedback, in contrast to the terminology used e.g. in \cite{Lam97},
where the term ``feedforward'' is used to indicate the position of
the displacement in the continuous beam.

Theoretically, the effects of the measurement on the state in the
optical pulse is described by eq.(\ref{eq:cond}), and the
feedback conditioned by the measurement result can be described
by a unitary displacement operator $\hat{D}(f {\vec{\beta}})$
acting on this conditional output state
(For a detailed discussion of the displacement operator,
see e.g. \cite{WM}).
In order to separate the measurement noise from the input
polarization, it is convenient to exchange the ordering of
the field operators $\hat{a}_H^\dagger$ and $\hat{a}_V^\dagger$
and the displacement operator $\hat{D}(f {\vec{\beta}})$
using the following relations,
\begin{eqnarray}
\label{eq:displace}
\hat{D}(f \beta_H, f \beta_V)\left( \hat{a}_H^\dagger +
f \beta_H^* \right)
&=&
\hat{a}_H^\dagger \hat{D}(f \beta_H, f \beta_V)
\nonumber \\
\hat{D}(f \beta_H, f \beta_V)\left( \hat{a}_V^\dagger +
f \beta_V^* \right)
&=&
\hat{a}_V^\dagger \hat{D}(f \beta_H, f \beta_V).
\end{eqnarray}
Comparison with eq.(\ref{eq:cond}) then shows that a special feedback
condition exists where the displacement can eliminate the
products of the input state amplitudes $c_{H/V}$ and the
measurement results $\beta_{H/V}^*$. Specifically,
the feedback compensated output state for a feedback factor
of $f_R = \sqrt{R/(1-R)}$ reads
\begin{eqnarray}
\label{eq:fb}
\lefteqn{
\sqrt{p({\vec{\beta}})} \;
\hat{D}( f_R \beta_H,  f_R \beta_V)
\mid \psi({\vec{\beta}}) \rangle_{a} =
}
\nonumber \\ &&
  \hspace{1cm} \frac{1}{\pi} \,
\mbox{e}^{-|\vec{\beta}|^2/2}
\sqrt{1-R} \left( c_H \hat{a}_H^\dagger +
c_V \hat{a}_V^\dagger \right)
\hat{D}( f_R \beta_H, f_R \beta_V)
\mid 0;0\rangle_{a}.
\end{eqnarray}
The output state at this special feedback condition
is therefore described by the action of the creation operator
of the original input photon on a coherent state.
As the analogy to cloning by photon bunching \cite{Irv04} suggests,
such an application of the single photon creation operator
to an otherwise random state describes an optimal cloning process.
For the single photon input, the unique feedback condition
$f_R=\sqrt{R/(1-R)}$ thus converts the beam splitter attenuation
into an optimal cloning process.

As first shown in \cite{Lam97}, the reason for the existence of the
optimal feedback condition can be explained in terms of the
quantum noise in the linear field amplification realized by the
setup shown in fig.\ref{fig1}. Specifically, the feedback condition
$f_R=\sqrt{R/(1-R)}$ exactly compensates the effects of the
vacuum noise entering at the beam splitter of reflectivity $R$,
leaving only the uncertainty limited noise caused by the simultaneous
measurement of both quadrature components.
Since the setup shown in fig. \ref{fig1} can also be used to
amplify and clone coherent states as demonstrated in \cite{And05},
it may be instructive to express the optimal feedback
condition in terms of the field gain that would be obtained for
such coherent input fields. If the input was a coherent state with
an average amplitude of ${\bf \alpha}$, the average measurement result
${\vec{\beta}}$ of the reflected light would be equal to
$\sqrt{R}\; {\bf \alpha}$ and the feedback would add
an average amplitude of $ f \sqrt{R}\; {\bf \alpha}$
to the transmitted amplitude of $\sqrt{1-R}\; {\bf \alpha}$,
for a total amplitude of $ g \; {\bf \alpha}$, where the
gain factor of the amplification is
$g=f \sqrt{R} + \sqrt{1-R}$. The special feedback condition
$f_R$ thus corresponds to a gain factor of
\begin{equation}
g_R = f_R \sqrt{R} + \sqrt{1-R} = 1/\sqrt{1-R}.
\end{equation}
Interestingly, this result indicates that optimal cloning is achieved
when the gain is exactly the inverse of the attenuation suffered by
the transmitted amplitude at the beam splitter.

\section{Output density matrix of the optimal cloning process}
To show that the effect of this beam splitter amplification
on a single photon input is indeed an optimal cloning process,
it is necessary to consider the output statistics averaged
over all measurement results ${\vec{\beta}}$. The density
operator of the output state is given by
\begin{eqnarray}
\label{eq:out}
\hat{\rho}(\mbox{out}) &=&
\int d^4{\vec{\beta}}\; p({\vec{\beta}})\;
\hat{D}( f_R \beta_H,  f_R \beta_V)
\mid \psi({\vec{\beta}}) \rangle \langle
\psi({\vec{\beta}}) \mid \hat{D}^\dagger( f_R \beta_H,  f_R \beta_V)
\nonumber \\
&=& (1-R) \Big(c_H \hat{a}_H^\dagger +
c_V \hat{a}_V^\dagger \Big) \; \hat{\eta}_R \;
\Big(c_H^* \hat{a}_H +
c_V^* \hat{a}_V \Big),
\end{eqnarray}
where the operator $\hat{\eta}_R$ is the density operator of a
thermal light field state with an average photon number of
$f_R^2=R/(1-R)$ in each mode. The cloning process is thus
described by the application of the input photon creation
operator to a completely unpolarized light field state.
It is now possible to separate $\hat{\rho}(\mbox{out})$ into
contributions with different output photon number $N$,
\begin{eqnarray}
\hat{\rho}(\mbox{out}) &=&
\sum_{N=1}^\infty P(N) \Big(c_H \hat{a}_H^\dagger +
c_V \hat{a}_V^\dagger \Big) \; \hat{C}_N \;
\Big(c_H^* \hat{a}_H +
c_V^* \hat{a}_V \Big),
\end{eqnarray}
where $P(N)$ is the probability of an $N$-photon output,
and $\hat{C}_N$ is the properly normalized operator of the
completely unpolarized $(N-1)$-photon state before the
application of the input photon creation operator,
\begin{eqnarray}
\label{eq:PN}
P(N) &=& \frac{(1-R)^3}{2 R} R^N N(N+1),
\\
\label{eq:CN}
\hat{C}_N &=& \frac{2}{N(N+1)} \sum_{n=1}^{N}
\mid n-1;N-n \rangle \langle n-1; N-n \mid.
\end{eqnarray}
The average number of clones can be controlled by varying
the reflectivity $R$ of the beam splitter, with high
reflectivities generating large numbers of clones and low
reflectivities generating only a few clones.
It is thus far easier to increase the number of clones than
in cloning methods relying on parametric downconversion,
where it is rather difficult to increase the parametric
gain \cite{Lam01}. For practical purposes, however, it may
be desirable to keep the cloning probabilities low, since
the quantum efficiency of photon detection is usually limited,
and the only way to ensure that the detection of $N$ photons
really corresponds to $N$ output photons is to keep the
probability of generating $N+1$ photons much lower than the
probability for $N$ photons. Eq. (\ref{eq:PN}) is therefore
essential for the optimal choice of $R$ in an experiment
with limited detector efficiencies.

Using the operator $\hat{C}_N$, it is now possible to
determine the output statistics of the $1 \to N$ photon
cloning process. The normalized density matrix
$\hat{\rho}_N$ of the $N$-photon output reads
\begin{equation}
\hat{\rho}_N = \Big(c_H \hat{a}_H^\dagger +
c_V \hat{a}_V^\dagger \Big) \; \hat{C}_N \;
\Big(c_H^* \hat{a}_H +
c_V^* \hat{a}_V \Big).
\end{equation}
This output is a mixture of photon number states with
$n$ photons in the correct input polarization and
$N-n$ photons in the opposite polarization.
The statistical weight of each state is determined
by the factor of $n$ introduced by the application
of the creation operator of the input photon to both
sides of the unpolarized operator $\hat{C}_N$.
The normalized probability distribution $P(n|N)$ of
the number of correctly polarized photons $n$ among $N$
output photons thus reads
\begin{equation}
P(n|N)=\frac{2n}{N(N+1)}.
\end{equation}
The fidelity of the $1 \to N$ photon
cloning process is then given by the ratio between the
average photon number in the input polarization and the
total output photon number,
\begin{equation}
F_{1 \to N} = \sum_{n=0}^N p(n|N) \frac{n}{N} =
\frac{2N+1}{3N}.
\end{equation}
This is the optimal fidelity for $1 \to N$ cloning \cite{Gis97}.
Thus, the coherent feedback setup shown in fig. \ref{fig1} is
indeed an optimal cloning machine.

The analysis above assumed precise homodyne detection and field
discplacements. In realistic implementations, it will be necessary
to take into account additional errors. In particular, nonunit
quantum efficiency and mode matching may introduce errors in the
linear amplification scheme, as discussed in some detail in
\cite{Oli06}. While a detailed analysis is beyond the scope
of the present paper, it may be worthwhile to consider the possible
effects of such errors on the $N$ photon polarization observed in
the output. Assuming that the errors are not polarization sensitive,
we may assume that they can be represented by a "white noise"
background of equal probability, as given by the density operator
$\hat{W}_N = \hat{C}_{N+1}$. The actual cloning fidelity achieved
will then be a weighted average of the optimal cloning fidelity and the
"white noise" fidelity of $1/2$.

\section{Information and noise in the coherent cloning process}
As the discussion above has shown, it is in principle possible
to realize optimal cloning of a single photon
polarization state by applying quantum measurements to the continuous
field variables. The field measurement projects the transmitted
field into a coherent superposition of vacuum and single photon
components as given by eq.(\ref{eq:cond}). By choosing an optimal feedback
gain of $g_R=1/\sqrt{1-R}$, this superposition of vacuum and single
photon component can be converted into a superposition of optimally
cloned $N$-photon outputs, as given by eq.(\ref{eq:fb}). Remarkably,
this manipulation of photon polarization states is achieved entirely
by continuous variable operations, making use of the correspondence
between the quantum coherence of the single photon state and the
(classical) optical coherence of the field.

It may also be worth noting that, due to the formal equivalence
of continuous variable teleportation errors and beam splitter
losses \cite{Hof01}, a closely related optimal cloning process
can be implemented by the continuous variable teleportation
of single photon states \cite{Pol99,Ide02a,Ide02b}.
In this case, the optimal gain condition depends on the squeezed state
entanglement, with low entanglement requiring a
correspondingly higher gain to achieve optimal cloning.
At a teleportation gain of $g=1$, the cloning process is
not optimal and the teleportation errors in the $N$ photon
outputs will be greater than the minimal cloning errors
\cite{Ide05}. In both cases, the essential feature of the
cloning process is that continuous variable measurements and
field displacements are used to clone the polarization states
of individual input photons.

Besides demonstrating the potential usefulness of continuous
variable operations for the processing of photon polarization qubits,
the other significant feature that distinguishes the present
cloning scheme from previous proposals for the cloning of photon
polarization is the use of projective
measurements to implement optimal cloning. It may be interesting
to note that the cloning process can be optimized despite the
partial loss of the original photon at the beam splitter.
Obviously, the loss of quantum information due to
the partial absorption of the input photon is balanced
by the usefulness of the classical measurement information
$\vec{\beta}$ for the generation of quantum clones. This
balance could also be used to explain the existence of an optimal gain:
at $g>1/\sqrt{1-R}$, the measurement information obtained
is not sufficient to optimize the fidelity of the high number
of clones generated, and at $g<1/\sqrt{1-R}$, the
measurement back action caused by the unnecessary precision
of the measurement introduces additional cloning errors.
The present cloning methods may thus provide some insights
into the relation between classical information and quantum
information in quantum cloning processes.

\section{Conclusions}
In conclusion, we have shown that an amplification of the
light field by coherent feedback of the reflection losses
at a beam splitter of reflectivity $R$ optimally clones
the polarization state of a single photon input if the
feedback induced gain is equal to $g_R=1/\sqrt{1-R}$.
This cloning method does not require any non-linear
optical elements and multiple clones are easy to obtain.
It may therefore be particularly useful for closing
the gap between output photon numbers of $N=2$ and
$N\to \infty$. By employing field measurements to
manipulate the polarization states of photons,
this cloning methods also illustrates the fundamental
relation between the continuous field variables and
the discrete photon number distributions of the
quantized light field. Photon cloning by coherent
feedback amplification thus shows how fundamental aspects
of the wave-particle dualism of light can be applied
to realize quantum information processes.
\ack
Part of this work has been supported by the JST-CREST
project on quantum information processing and the
Grant-in-Aid program of the Japanese Society for the
Promotion of Science.

\end{document}